\documentclass[aps,pre,twocolumn,groupedaddress]{revtex4-1}
\usepackage{amsmath} % AMS Math Package
\newcommand{\avg}[1]{\left< #1 \right>} % for average
\newcommand{\dZ}[0]{\Delta Z} 
\newcommand{\dZi}[0]{\Delta Z_\text{initial}} 
\newcommand{\muBm}[0]{{\mu_{B_-}}}
\newcommand{\muBp}[0]{{\mu_{B_+}}}
\newcommand{\muGp}[0]{{\mu_{G_+}}}
\newcommand{\eq}[1]{\begin{align} #1 \end{align}}

\usepackage{graphics} % or graphicx 
\usepackage{hyperref} %clickable references.
\hypersetup{
    colorlinks,
    citecolor=blue,
    filecolor=black,
    linkcolor=blue,
    urlcolor=blue
}

\usepackage{epstopdf}
\usepackage{epsfig}
\usepackage{tikz}

\begin{document}

%\preprint{}

%\title{When do Jammed Particulate Packings have a Linear Regime?}
%\title{The validity of linear response in jammed particulate packings}
\title{Tuning by pruning: exploiting disorder for global response and the principle of bond-level independence}


\author{Carl P. Goodrich}
\email[]{cpgoodri@sas.upenn.edu}
%\thanks{}
%\affiliation{University of Pennsylvania}

\author{Andrea J. Liu}
%\email[]{ajliu@physics.upenn.edu}
%\thanks{}
\affiliation{Department of Physics, University of Pennsylvania, Philadelphia, Pennsylvania 19104, USA}

\author{Sidney R. Nagel}
%\thanks{}
\affiliation{James Franck Institute, The University of Chicago, Chicago, Illinois 60637, USA}

\date{\today}

\begin{abstract}
We exploit the intrinsic difference between disordered and crystalline solids to create systems with unusual and exquisitely tuned mechanical properties. To demonstrate the power of this approach, we design materials that are either virtually incompressible or completely auxetic. Disordered networks can be efficiently driven to these extreme limits by removing a very small fraction of bonds via a selected-bond removal procedure that is both simple and experimentally relevant. The procedure relies on the nearly complete absence of any correlation between the contributions of an individual bond to different elastic moduli.  A new principle unique to disordered solids underlies this lack of correlation: independence of bond-level response.
\end{abstract}

%\keywords{}

\maketitle

The properties of amorphous solids are essentially and qualitatively different from those of simple crystals~\cite{Goodrich:2014fl}.  In a crystal, identical unit cells are interminably and symmetrically repeated, ensuring that all cells make identical contributions to the solid's global response to an external perturbation~\cite{Ashcroft:1976ud,kittel2004introduction}.  Unless a crystal's unit cell is very complicated, all particles or inter-particle bonds contribute nearly equally to any global quantity, so that each bond plays a similar role in determining the physical properties of the solid. For example, removing a bond in an ordered array or network decreases the overall elastic strength of the system, but in such a way that the resistance to shear and the resistance to compression drop in tandem~\cite{Feng:1985vr} so that their ratio is nearly unaffected.
%roughly proportionally to one another~\cite{Feng:1985vr,Ellenbroek:2009to}. 
Disordered materials are not similarly constrained. We will show that as a consequence, one can exploit disorder to achieve a unique, varied, textured and tunable global response.

A tunable global response is a corollary to a new principle that emerges for disordered matter: independence of bond-level response.  This independence refers not only to the dearth of strong correlations between the response of  different bonds, but also, and more importantly, to the response of any specific bond to different external perturbations.  We will demonstrate this by constructing selected-bond-removal networks, where individual bonds, or springs, are successively removed to drive the overall system into different regimes of behavior, characterized by ratios of different mechanical responses.  
Starting from the same initial network, we can remove as few as 2\% of the bonds to produce a network with a ratio of the shear to bulk modulus, $G/B$, that is either nearly zero (incompressible limit) or nearly infinite (maximally auxetic~\cite{Greaves:2011ku}) merely by removing different sets of bonds. Moreover, by using different algorithms or starting with different configurations, we find that the region within which the bonds are removed can be confined to strips of controllable size, ranging from a few bond lengths to the size of the entire sample. This has the practical consequence that one can achieve precise spatial control in tuning  properties of the material from region to region within the network--as is needed for creating origami~\cite{Witten:2007cq,Mahadevan:2005hr} or kirigami~\cite{Castle:2014jg} materials. 

We construct networks numerically by starting with a configuration of particles produced by a standard jamming algorithm~\cite{OHern:2003vq,Liu:2010jx}.  
We place $N$ soft repulsive particles at random in a box of linear size $L$ and minimize the total energy until there is force balance on each particle.  
%We place $N$ soft repulsive particles at random in a box of linear size $L$ and quench the total energy to a local minimum so that there is force balance on each particle. This results in a disordered packing that is well characterized~\cite{OHern:2003vq,Liu:2010jx}.
We work in either two or three dimensions and start with a packing fraction, $\phi$, that is above the jamming density.  After minimizing the energy of a configuration, we create a network by replacing each pair of interacting particles with an unstretched spring of unit stiffness between nodes at the particle centers~\cite{Wyart:2005jna}. We characterize the network by the excess coordination number $\dZ \equiv Z - Z_\text{iso}$, where $Z$ is the average number of bonds at each node and $Z_\text{iso} \equiv 2d - 2d/N$ is the minimum for a system to maintain rigidity in $d$ dimensions~\cite{Goodrich:2012ck}.

\begin{figure}[htpb]
	\centering
	\includegraphics[width=\linewidth]{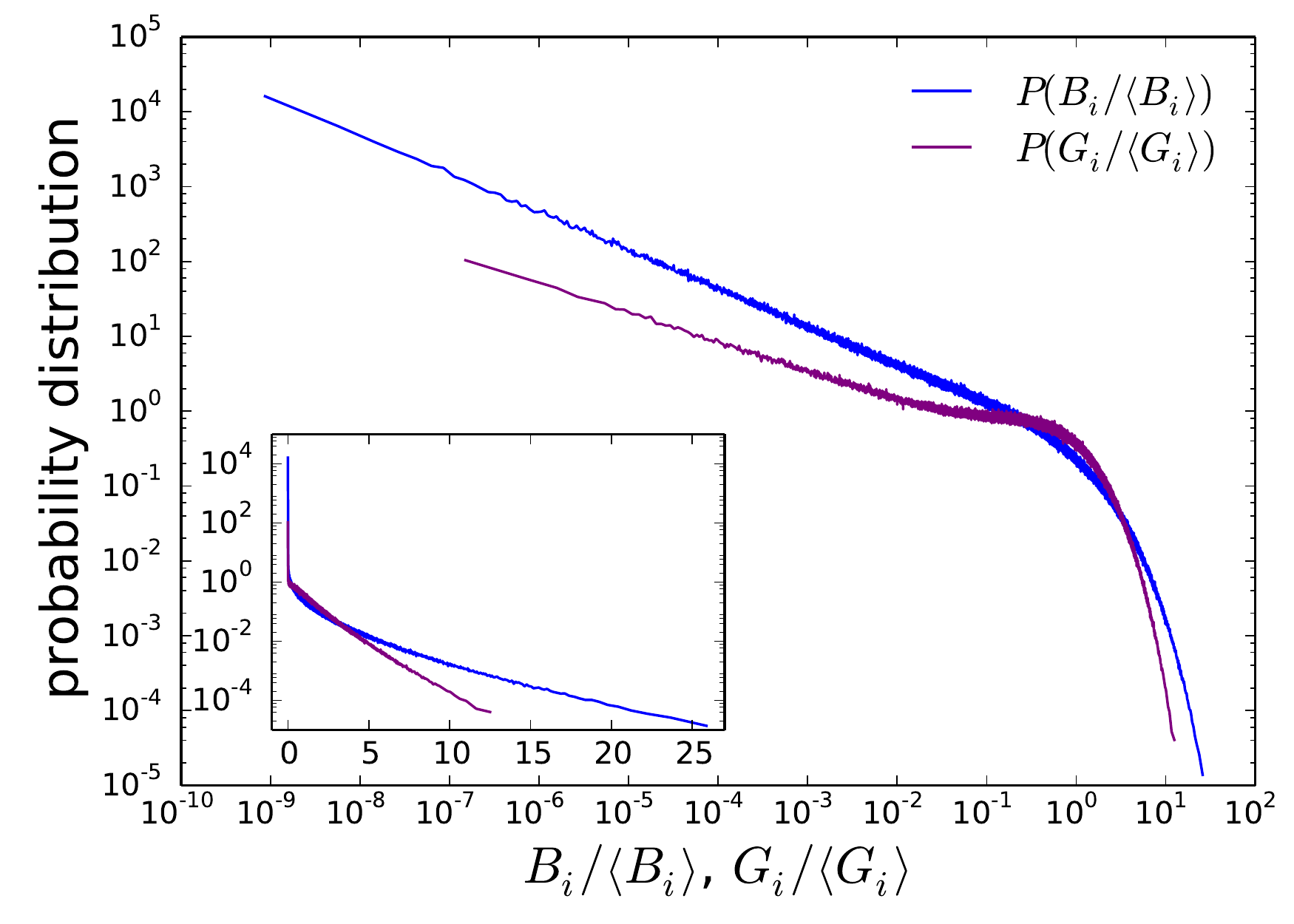}
	\caption{\label{fig:Ri_distributions_3d}Bond-level response. Distribution on a log-log scale (inset: log-linear scale) of the contribution of each bond to the macroscopic bulk and shear moduli, $B_i$ and $G_i$, for $3d$ networks with $\dZi \approx 0.127$. Here $i$ indexes bonds. 
	%The distributions are normalized by their means. 
	At low $B_i$ or $G_i$, the distributions follow power-laws with exponents $-0.51$ and $-0.38$, respectively.  At high values, the distributions decay over a range that is broad compared to their means, $\avg{B_i}$ and $\avg{G_i}$.} 
\end{figure}

For each network, we use linear response to calculate the contribution $B_i$ of each bond $i$ to the bulk modulus, $B=\sum_i B_i$ (see Appendix for details). 
The distribution of $B_i$ in three dimensions is shown in blue in Fig.~\ref{fig:Ri_distributions_3d}, where data are averaged over 500 configurations, each with approximately 4000 nodes and an initial excess coordination number $\dZi \approx 0.127$ (corresponding to a total number of bonds that is about 2\% above the minimum needed for rigidity).

Similarly, we can start with the same initial network and calculate $G_i$, the contribution of each bond to the angle-averaged shear modulus, $G=\sum_i G_i$.  (A finite system is not completely isotropic, so the shear modulus varies with direction~\cite{DagoisBohy:2012dh}; we calculate the angle-averaged shear modulus, which approaches the isotropic shear modulus in the infinite system size limit~\cite{Goodrich:2014iu}.)   The resulting distribution for $G_i$ is shown in purple in Fig.~\ref{fig:Ri_distributions_3d}.  Note that the distributions of the bond contributions to $B$ and $G$ are continuous, very broad, and non-zero in the limit $B_i,G_i \rightarrow 0$.  That is, some bonds have nearly zero contribution to the bulk or shear modulus while others contribute disproportionately.  For both $B$ and $G$, the distribution decays as a power law at low values of $B_i$ or $G_i$. These power laws are terminated above $\avg{B_i}$ and $\avg{G_i}$ by approximately exponential cut-offs. In comparison, the distributions for a perfect crystal would be composed of discrete delta functions.

\begin{figure}[htpb]
	\centering
	\includegraphics[width=0.9\linewidth]{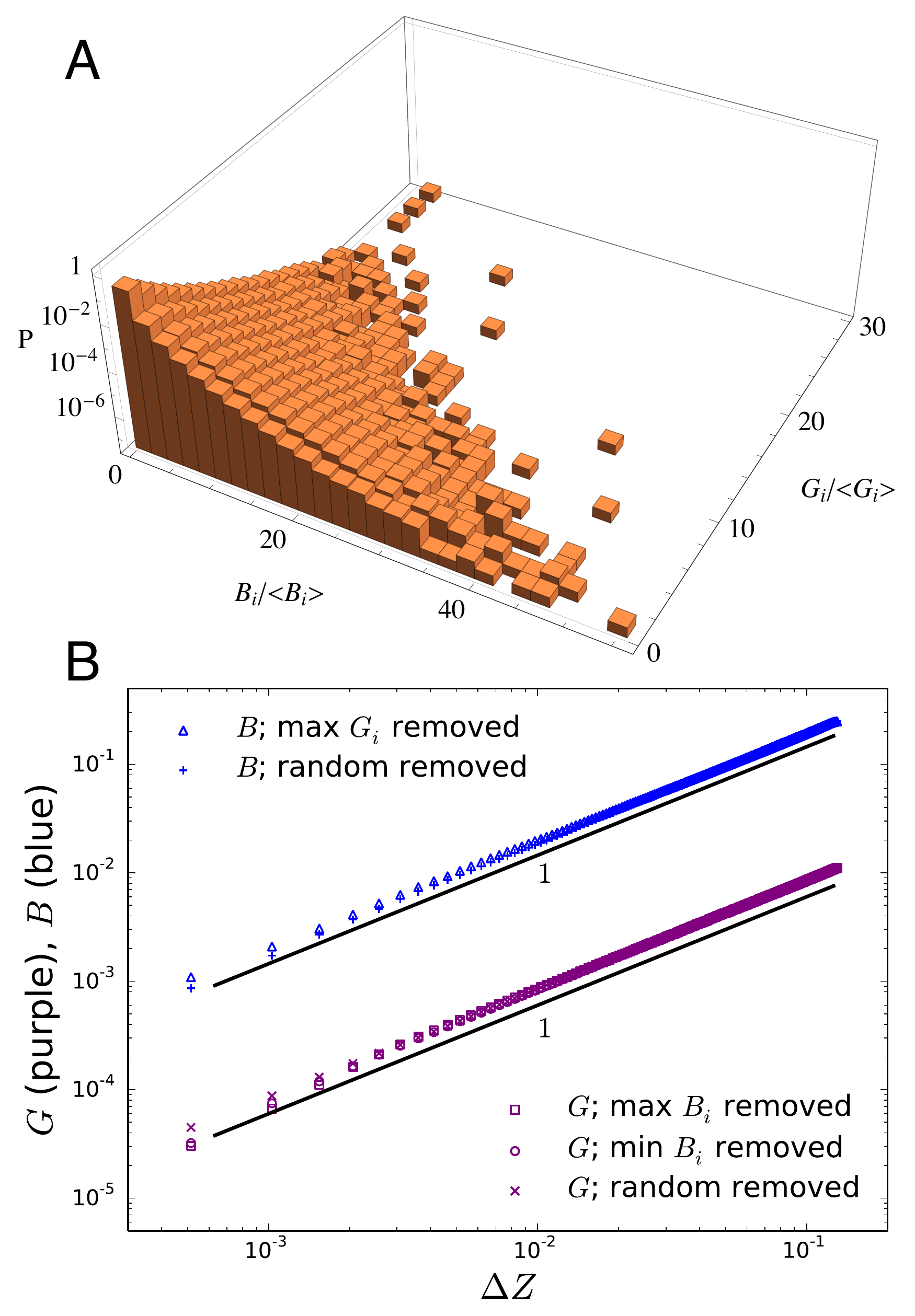} 
	\caption{\label{fig:bond_level_independence}Independence of bond-level response. ({\bf A}) Joint probability distribution of $B_i$ and $G_i$ for $3d$ networks with $\dZi \approx 0.127$. There is little apparent correlation between the response to compression ($B_i$) and to shear ($G_i$) for a given bond $i$.  ({\bf B}) The value of $G$ when bonds with the largest (purple squares) and smallest (purple circles) $B_i$ are removed is nearly indistinguishable from $G$ when bonds are removed at random (purple crosses). Similarly, $B$ is very similar whether bonds with the largest $G_i$ (blue triangles) are removed or bonds are removed at random (blue pluses).}
\end{figure}

\begin{figure}[htpb]
	\centering
	\includegraphics[width=0.9\linewidth]{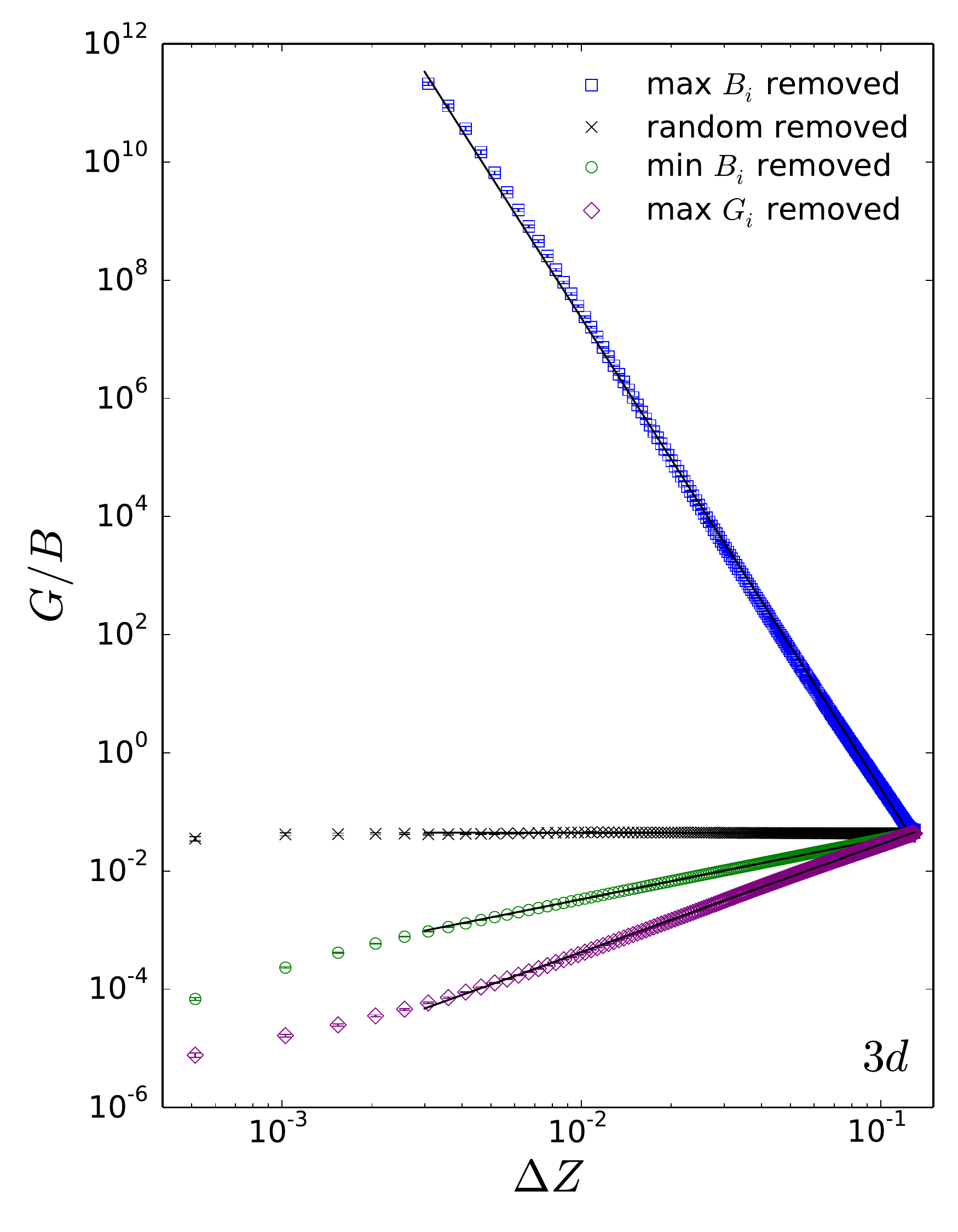} 
	\caption{\label{fig:Global_response}Tuning global response in three dimensions. The ratio of shear to bulk modulus, $G/B$, for  four pruning algorithms. Error bars (included) are smaller than the symbols. Lines are fits to the data over the indicated range and have slopes, from top to bottom, of -7.96, -0.01, 1.01, and 1.82. Starting with the same initial conditions, we can tune global response by 16 orders of magnitude by pruning of order 2\% of the bonds.}
\end{figure}

\begin{figure}[htpb]
	\centering
	\includegraphics[width=0.9\linewidth]{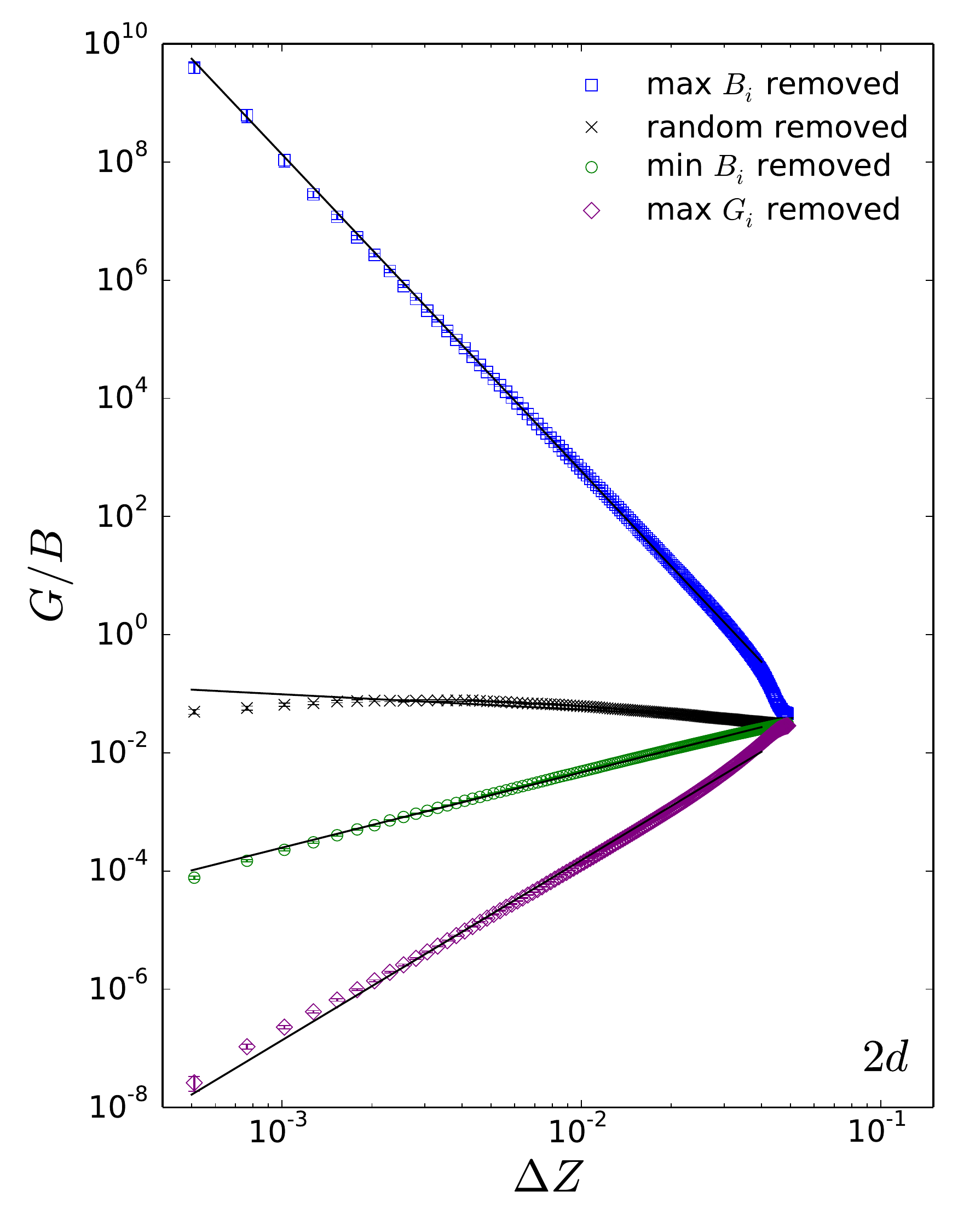} 
	\caption{\label{fig:Global_response_2d}Tuning global response in two dimensions. The ratio of shear to bulk modulus, $G/B$, for  four pruning algorithms. Error bars (included) are smaller than the symbols. Lines are fits to the data over the indicated range and have slopes, from top to bottom, of -5.36, -0.26, 1.27, and 3.05. Starting with the same initial conditions, we can tune global response by 17 orders of magnitude by pruning of order 1\% of the bonds.}
\end{figure}

We next ask if there is a correlation between how an individual bond responds to shear and how it responds to compression.  Do bonds with a large contribution to the bulk modulus also have a proportionately large contribution to the shear modulus?  Fig.~\ref{fig:bond_level_independence}a shows the joint probability distribution $P(B_i,G_i)$. 
There is a nearly vanishing (but not identically zero) correlation between how individual bonds respond to shear and how they respond to compression. This is qualitatively different from what one would find for a simple crystal. 
Thus, Fig.~\ref{fig:bond_level_independence}a illustrates a previously-unrecognized property that is very well obeyed by disordered networks: independence of bond-level response.  

This new property suggests that one can tailor the behavior of the network by selectively removing (pruning) those bonds that contribute more or less than the average to one of the moduli.  By so doing, one can decrease one modulus with respect to the other.  

First, we consider the known case of \emph{rigidity percolation}~\cite{Feng:1985vr,Ellenbroek:2009to,Ellenbroek:2014uh}, where a bond is picked at random and removed. This pruning is repeated until the system becomes unstable at $\dZ = 0$. We have implemented a slight variation to this procedure: at each step, a bond is removed only if each node connected to this bond has at least $d+1$ remaining bonds in $d$ dimensions.  This is the condition for local stability of a particle in the original jammed packing~\cite{Levine2001}.
As the excess coordination number decreases, the bulk and shear moduli vanish together, so that $G \sim B \sim \dZ$~\cite{Feng:1985vr,Ellenbroek:2009to,Ellenbroek:2014uh} (see Fig.~\ref{fig:bond_level_independence}b).
Therefore, as shown in Fig.~\ref{fig:Global_response}, the ratio $G/B$ is independent of $\dZ$.

We now implement the idea of \emph{selected}-bond removal in a variety of ways.   
First we remove the bond with the smallest $B_i$, namely the weakest contribution to the bulk modulus (provided, as above, that each node connected to this bond has at least $d+1$ remaining bonds). Since the distribution $P(B_i)$ is continuous and nonzero as $B_i \rightarrow 0$, the bond removal has almost no effect on the bulk modulus. However, since there is little correlation between the contribution of each bond to the bulk and shear moduli, there is a much larger effect on the shear modulus. 
The contributions $B_i$ and $G_i$ of the remaining bonds to the moduli are then recalculated and the procedure is repeated to remove the bond with the smallest $B_i$.  Figure~\ref{fig:bond_level_independence}b shows that when bonds with the smallest $B_i$ are successively removed, the \emph{shear} modulus linearly proportional to $\Delta Z$. Furthermore, it is quantitatively identical, within numerical precision, to when bonds are removed at random. 
The ability to alter the scaling of the bulk modulus without affecting the scaling of the shear modulus is a clear demonstration that the principle of independence of bond-level response allows for very precise tuning of global properties. 
%The ability to target bonds that contribute the least to the bulk modulus without affecting the shear modulus is a clear demonstration that the principle of independence of bond-level response allows for very precise tuning of global properties. 

Since removing bonds with the smallest $B_i$ has little effect on the bulk modulus, we would expect $G/B \rightarrow 0$ as $\dZ \rightarrow 0$. As shown in Fig.~\ref{fig:Global_response}, we find that $G/B \sim \dZ^\muBm$, with $\muBm = 1.01 \pm 0.01$. This behavior is identical to the scaling found in the original jammed sphere packings, where $\dZ$ is lowered by decompressing the system. %rather than targeting specific bonds for removal. 
%In decompressing a jammed packing, the contacts most likely to disappear are those with the least interparticle overlap, which contribute minimally to the bulk modulus.  This provides theoretical insight into why jamming has anomalous $G/B$ behavior.
In decompressing a jammed packing, this suggests that the contacts most likely to disappear are those which contribute minimally to the bulk modulus, providing theoretical insight into why jamming has anomalous $G/B$ behavior.

We can drive the same initial network to the opposite limit, $G/B \rightarrow \infty$, by successively removing bonds with the \emph{largest} contribution to $B$. 
As before, independence of bond-level response predicts that the shear modulus will again decrease linearly with $\dZ$, as we indeed find (see Fig.~\ref{fig:bond_level_independence}b). However, the bulk modulus will decrease more quickly, as prescribed by the high $B_i$ tail of the distribution, suggesting that the ratio $G/B$ should \emph{increase}. 
%The removal of the bond leads to redistribution of the load on the remaining bonds; we recalculate $G_i$ and $B_i$ for the remaining bonds and again remove the bond with the largest value of $B_i$.
The result of this successive bond-removal algorithm is shown by the blue squares in Fig.~\ref{fig:Global_response}.  We find that $G/B \sim \dZ^{\muBp}$, where $\muBp = -7.96 \pm 0.01$.  Thus, the increase in $G/B$ occurs with a \emph{much} steeper power law than the decrease of $G/B$ when the bond with the smallest contribution to $B$ is removed.  This power law implies that the distribution $P(B_i/\left<B_i\right>)$ evolves as bond pruning proceeds.

The algorithms mentioned above can be extended in a number of ways. For example, 
one can remove the bond with the largest contribution to the shear modulus to drive $G/B$ towards zero. In this case, independence of bond-level response implies that the bulk modulus would respond as if bonds were removed randomly, so that $B \sim \dZ$ (see Fig.~\ref{fig:bond_level_independence}b). However, the shear modulus decreases more rapidly; we find $G/B \sim \dZ^{\muGp}$, where $\muGp = 1.82 \pm 0.01$ (purple diamonds in Fig.~\ref{fig:Global_response}).

We can also tune two-dimensional networks with equal ease. We construct spring networks in two dimensions with approximately 8000 nodes and an initial coordination number of $\dZi \approx 0.047$, which is about 1\% above the minimum needed for rigidity.  
Figure~\ref{fig:Global_response_2d} shows $G/B$ as bonds are pruned towards $\dZ \rightarrow 0$ for the same four selected-bond removal algorithms  as in Fig.~\ref{fig:Global_response}.
% When bonds are randomly pruned (rigidity percolation), $G/B$ is roughly constant. The slight curvature we measure may be a consequence of the fact that fluctuations in connectivity are suppressed in three dimensions but are marginally critical in two dimensions~({\it 9}). 
 When bonds with the smallest $B_i$ are removed, we find that $G/B \sim \dZ^\muBm$ with $\muBm = 1.27 \pm 0.01$. This is close to the behavior known for jammed packings ($G/B \sim \dZ^1$), though it is certainly not as clean as in three dimensions. %, where $\muBm = 1.01 \pm 0.01$. 
When we prune bonds that resist compression the most (largest $B_i$), we find that $G/B \sim \dZ^\muBp$, where $\muBp = -5.36 \pm 0.01$. 
%While this is not as drastic as in three dimensions, where the power law is nearly $-8$, this is still \emph{extremely} steep. % with $G$ nearly $10^{10}$ times larger than $B$ at the smallest $\dZ$. 
At the smallest $\dZ$, $G/B \sim 10^{10}$.
Finally, when bonds with the largest $G_i$ are removed we find that $G/B \sim \dZ^\muGp$, with $\muGp = 3.05 \pm 0.01$.%, which vanishes even faster than in three dimensions (where $\muGp = 1.82 \pm 0.01$). This may be due to the fact that there are fewer independent shear directions in two dimensions (2) than in three dimensions (5), making it easier for the system to soften in all directions simultaneously. %However, further investigation is needed to confirm this interpretation. 
Although $G/B$ diverges/vanishes with slightly different power laws in two dimensions, the overall effect is no less drastic.  %As discussed in the main text, underlying this remarkable sensitivity is the principle of bond-level independence. 

%The limit $G/B \rightarrow 0$ corresponds to the incompressible limit of a solid where the Poisson ratio, $\nu= (d - 2 G/B)/ [d (d-1)+ 2 G/B ]$ in $d$ dimensions, reaches its maximum value of  $\nu=+1$ (in $2d$) or $+1/2$ (in $3d$).   The limit $G/B \rightarrow \infty$ corresponds to the auxetic limit where the Poisson ratio reaches its minimum value of $\nu=-1$.  By using these different pruning algorithms, we can tailor networks to have any Poisson ratio between these two limits.  This ability provides great flexibility in the design of network materials.  

Note that our procedures are remarkably efficient in tuning $G/B$.  Figures~\ref{fig:Global_response} and \ref{fig:Global_response_2d} show that by removing less than 2\% of the bonds in three-dimensional networks we can obtain a difference of more than 16 orders of magnitude in the tuned value of $G/B$, depending on which bonds we prune.  In two dimensions, pruning is similarly efficient; starting with the same initial configuration we are able to obtain differences in $G/B$ that span over 17 orders of magnitude by pruning only $\sim 1\%$ of the bonds.We also note that our bond-cutting procedures do not create any zero-frequency vibrational modes in the system, which would herald an instability in the structure. % Furthermore, Fig.~S1 shows that the response of $2d$ networks can be tuned just as dramatically~\cite{supplement}.

The limit $G/B \rightarrow 0$ corresponds to the incompressible limit of a solid where the Poisson ratio, $\nu= (d - 2 G/B)/ [d (d-1)+ 2 G/B ]$ in $d$ dimensions, reaches its maximum value of  $\nu=+1$ (in $2d$) or $+1/2$ (in $3d$).   The limit $G/B \rightarrow \infty$ corresponds to the auxetic limit where the Poisson ratio reaches its minimum value of $\nu=-1$.  By using these different pruning algorithms, we can tailor networks to have any Poisson ratio between these two limits.  This ability provides great flexibility in the design of network materials.

We turn now to spatial correlations between cut bonds.  Driscoll {\it et al.}~\cite{Driscoll:2015vf} have conducted numerical simulations in which they removed bonds with the {\it largest} strain under uniaxial or isotropic compression or shear.  They showed that  the cut bonds form a damage zone whose width increases and diverges as the initial excess coordination number, $\dZi \rightarrow 0$; for sufficiently small $\dZi$, the pruned bonds are homogeneously distributed throughout the entire system. Outside this zone, they found that the network is essentially unaffected.

When pruning bonds with the {\it largest} contribution to $B$ or $G$, all the data presented thus far are for systems with a sufficiently small $\dZi$ so that the distribution of the cut bonds appears homogeneous.  
%Thus far, for all the data in systems where the {\it largest} contribution to $B$ or $G$ is removed, we have started with a sufficiently small value of $\dZi$ so that the distribution of the cut bonds appears homogeneous.  
In our simulations with large $\dZi$, where the damage zone is smaller than the size of our system, we find that $G/B$ still diverges/vanishes, but does so when $\Delta Z >0$. When we remove the bond with the {\it smallest} contribution to $B$ or $G$, the bonds are initially removed homogeneously throughout the system, independent of $\dZi$. The existence of tunable strong spatial correlations in the cut bonds, as found by Driscoll {\it et al.}~\cite{Driscoll:2015vf}, allows one to create textured materials spatially varying mechanical properties. 
%with a variety of textures.  
One region may be highly incompressible while a nearby region may be highly auxetic.  This offers a great variety in the mechanical response of these networks.

% \begin{figure}[htpb]
% \centering
% \includegraphics[width=0.4\linewidth]{Fig4_a.png} \\
% \includegraphics[width=0.4\linewidth]{Fig4_b.png} 
% \caption{\label{fig:cut_bonds_image}Spatial correlations of pruned bonds. Typical $2d$ bond networks with an initial excess coordination number of ({\bf A}) $\dZi = 1.044$ and ({\bf B}) $\dZi = 0.047$. 
%At each pruning step, the bond with the largest $B_i$ is removed. When all the thick highlighted bonds are removed, both systems have $G/B \approx 10^6$. The color of the thick bonds indicates the order in which they are pruned on a red to blue scale, with red bonds removed first.}
% \end{figure}

For many materials~\cite{Greaves:2011ku} the Poisson ratio decreases with increased connectivity of the constituent particles and increases with packing density.  We note that neither of these correlations hold for the algorithms we have introduced for tuning the Poisson ratio (or ratio of shear and bulk moduli).  We can reach $G/B \rightarrow \infty$ (minimum Poisson ratio)  or $G/B \rightarrow 0$ (maximum Poisson ratio) by removing the same number of bonds from the same starting configuration. Neither the overall connectivity nor the overall density is different in the two final states.  Thus, our procedures for producing tunable Poisson ratio materials are fundamentally different from correlations considered in the literature.  

We have presented a number of ways of tuning $G/B$.  Our results suggest that these ideas may be extended to other global properties ({\it e.g.}, thermal expansion or electrical response~\cite{DEARCANGELIS:1985uh,DEARCANGELIS:1986vk}) where the response can be written in terms of sums over bond contributions. As long as there is independence of bond-level response, one should be able to tune the ratio of global properties by using the same protocol of removing bonds that are especially susceptible (or especially unsusceptible) to a given global perturbation.  

Our results demonstrate that disordered networks provide particularly elegant opportunities for constructing mechanical metamaterials with tunable, flexible and spatially textured response.  However, the algorithms we have presented may not be restricted to artificially constructed materials. 
For example, compressing a network composed of springs that fail when stressed past a given threshold would result in the same network as removing springs with the largest $B_i$, provided that the threshold is sufficiently small. 
It is also not beyond imagination that one could selectively break bonds at the nano-scale level in response to global perturbations in complex solids.  Indeed, biology appears to be able to target structures in networks that are under particularly high stress and to enhance their strength (such as in  trabecular bone~\cite{Keyak:2013ga}). Alternatively, there may be mechanisms to buckle or sever strongly stressed fibers (such as in actin networks~\cite{Lenz:2012df}).  It is interesting to ask if such selective repair or destruction of biological structures changes 
ratios of different mechanical responses such as the Poisson ratio.

\appendix
\section{Calculation of bond-level elastic response}
We consider networks of nodes connected by unstretched central-force springs with stiffness $k=1$.  Let $\vec{\delta r}_i$ be the total strain on bond $i$ when the system is deformed according to some strain tensor $\epsilon_{\alpha\beta}$. The change in energy of the network is then given to lowest order by
\eq{	\Delta E = \sum_{i} k_i \delta r_{i,\parallel}^2, \label{eq:energy_change}}
where $\delta r_{i,\parallel}$ is the component of $\vec {\delta r}_i$ that is parallel to the bond direction.  Thus, the bond that contributes the most (least) to the response to a given boundary deformation is the one with the largest (smallest) $\delta r_{i,\parallel}^2$. To remove the bond that contributes the most to the bulk modulus, for example, one would remove the bond with the largest $\delta r_{i,\parallel}^2$ under compression.  This procedure can be implemented in either a simulation or an experiment.  

In practice, for our computations, we use linear algebra to calculate the response of each bond more efficiently, as follows. 
The bulk elasticity of a system is described to linear order by the elastic modulus tensor $c_{\alpha\beta\gamma\delta}$, so that if the system is distorted by the symmetric strain tensor $\epsilon_{\alpha\beta}$, the change in energy is given to leading order by
\eq{	\Delta E/V = \frac 12 \epsilon_{\alpha\beta}c_{\alpha\beta\gamma\delta}\epsilon_{\gamma\delta}, }
where $V$ is the volume of the system.
In general, there are 6 (21) independent components of the elastic modulus tensor in two (three) dimensions, but in the isotropic limit this reduces to just the bulk modulus $B$ and the shear modulus $G$. 

The components of $c_{\alpha\beta\gamma\delta}$ are calculated from the change in energy of the system under various boundary deformations using Eq.~\ref{eq:energy_change}. %$\vec {\delta r}_i$ is composed of an affine strain and a non-affine response. 
The strain $\vec{\delta r}_i$ can be decomposed into two distinct parts. First there is an affine strain set directly by the strain tensor. However, this results in a nonzero net force, $\vec f_m$, on each node $m$, leading to a secondary non-affine response. This non-affine response is calculated by solving the following system of equations
\eq{	\mathcal{M}_{mn} \vec u^\text{NA}_{m} = \vec f_n,}
where %the indices $m$ and $n$ run over the $Nd$ degrees of freedom associated with node displacements, 
$\mathcal{M}_{mn}$ is the Hessian matrix and $\vec u^\text{NA}_m$ is the non-affine displacement of each node. The total strain $\vec {\delta r}_i$ of bond $i$ is calculated from the sum of the affine and non-affine displacements of the two nodes that the bond connects.
Since $\Delta E$ can be written as a sum over bonds, so too can the elastic modulus tensor:
\eq{	c_{\alpha\beta\gamma\delta} = \sum_i c_{i,\alpha\beta\gamma\delta}. }
Under the deformation $\epsilon_{\alpha\beta}$, the change in energy of bond $i$ is
\eq{	\Delta E_i = \frac 12 \epsilon_{\alpha\beta}c_{i,\alpha\beta\gamma\delta}\epsilon_{\gamma\delta}. }
$c_{i,\alpha\beta\gamma\delta}$ thus completely describes the bond-level elastic response for bond $i$, and can be used to calculate the quantities $B_i$ and $G_i$ considered in the main text. 

The global bulk and shear moduli are linear combinations of the components of the elastic modulus tensor. In two dimensions, they are
\eq{
	B &= \tfrac 14 \left( c_{xxxx} + c_{yyyy} + 2c_{xxyy}\right) \\
	G &= \tfrac 18 \left( 4c_{xyxy} + c_{xxxx} + c_{yyyy} - 2c_{xxyy}\right),
}
while in three dimensions they are
\eq{
	B ={}& \tfrac 19 \left( c_{xxxx} + c_{yyyy} + c_{zzzz} + 2c_{yyzz} + 2c_{xxzz} + 2c_{xxyy}\right) \\
	G ={}& \tfrac 1{15} \left( 3c_{yzyz} + 3c_{xzxz} + 3c_{xyxy} \right. \nonumber \\
	& \left. +\, c_{xxxx} + c_{yyyy} + c_{zzzz} - c_{yyzz} - c_{xxzz} - c_{xxyy}\right).
}
Finite disordered systems are never perfectly isotropic, so the shear modulus always has some dependence on the angle of shear. The above expressions for $G$ represent the angle-averaged shear modulus, which reduces to the shear modulus in the isotropic limit of infinite system size. We calculate the contribution of bond $i$ to the bulk and shear moduli in exactly the same way:
\eq{
	B_i &= \tfrac 14 \left( c_{i,xxxx} + c_{i,yyyy} + 2c_{i,xxyy}\right) \\
	G_i &= \tfrac 18 \left( 4c_{i,xyxy} + c_{i,xxxx} + c_{i,yyyy} - 2c_{i,xxyy}\right),
}
in two dimensions, and
\eq{
	B_i ={}& \tfrac 19 \left( c_{i,xxxx} + c_{i,yyyy} + c_{i,zzzz} + 2c_{i,yyzz} + 2c_{i,xxzz} + 2c_{i,xxyy}\right) \\
	G_i ={}& \tfrac 1{15} \left( 3c_{i,yzyz} + 3c_{i,xzxz} + 3c_{i,xyxy} \right. \nonumber \\
	& \left. +\, c_{i,xxxx} + c_{i,yyyy} + c_{i,zzzz} - c_{i,yyzz} - c_{i,xxzz} - c_{i,xxyy}\right)
}
in three dimensions. 

\begin{acknowledgments}
We thank Bryan Chen, Michelle Driscoll, Heinrich Jaeger and Vincenzo Vitelli for important discussions. This research was supported by the US Department of Energy, Office of Basic Energy Sciences, Division of Materials Sciences and Engineering under Awards DE-FG02-05ER46199 (A.J.L., C.P.G.) and DE-FG02-03ER46088 (S.R.N.). This work was partially supported by a grant from the Simons Foundation (\#305547 to A.J.L.).
\end{acknowledgments}

%\bibliography{papers2_bibtex.bib,extrabib.bib}

\begin{thebibliography}{22}%
\makeatletter
\providecommand \@ifxundefined [1]{%
 \@ifx{#1\undefined}
}%
\providecommand \@ifnum [1]{%
 \ifnum #1\expandafter \@firstoftwo
 \else \expandafter \@secondoftwo
 \fi
}%
\providecommand \@ifx [1]{%
 \ifx #1\expandafter \@firstoftwo
 \else \expandafter \@secondoftwo
 \fi
}%
\providecommand \natexlab [1]{#1}%
\providecommand \enquote  [1]{``#1''}%
\providecommand \bibnamefont  [1]{#1}%
\providecommand \bibfnamefont [1]{#1}%
\providecommand \citenamefont [1]{#1}%
\providecommand \href@noop [0]{\@secondoftwo}%
\providecommand \href [0]{\begingroup \@sanitize@url \@href}%
\providecommand \@href[1]{\@@startlink{#1}\@@href}%
\providecommand \@@href[1]{\endgroup#1\@@endlink}%
\providecommand \@sanitize@url [0]{\catcode `\\12\catcode `\$12\catcode
  `\&12\catcode `\#12\catcode `\^12\catcode `\_12\catcode `\%12\relax}%
\providecommand \@@startlink[1]{}%
\providecommand \@@endlink[0]{}%
\providecommand \url  [0]{\begingroup\@sanitize@url \@url }%
\providecommand \@url [1]{\endgroup\@href {#1}{\urlprefix }}%
\providecommand \urlprefix  [0]{URL }%
\providecommand \Eprint [0]{\href }%
\providecommand \doibase [0]{http://dx.doi.org/}%
\providecommand \selectlanguage [0]{\@gobble}%
\providecommand \bibinfo  [0]{\@secondoftwo}%
\providecommand \bibfield  [0]{\@secondoftwo}%
\providecommand \translation [1]{[#1]}%
\providecommand \BibitemOpen [0]{}%
\providecommand \bibitemStop [0]{}%
\providecommand \bibitemNoStop [0]{.\EOS\space}%
\providecommand \EOS [0]{\spacefactor3000\relax}%
\providecommand \BibitemShut  [1]{\csname bibitem#1\endcsname}%
\let\auto@bib@innerbib\@empty
%</preamble>
\bibitem [{\citenamefont {Goodrich}\ \emph
  {et~al.}(2014{\natexlab{a}})\citenamefont {Goodrich}, \citenamefont {Liu},\
  and\ \citenamefont {Nagel}}]{Goodrich:2014fl}%
  \BibitemOpen
  \bibfield  {author} {\bibinfo {author} {\bibfnamefont {C.~P.}\ \bibnamefont
  {Goodrich}}, \bibinfo {author} {\bibfnamefont {A.~J.}\ \bibnamefont {Liu}}, \
  and\ \bibinfo {author} {\bibfnamefont {S.~R.}\ \bibnamefont {Nagel}},\
  }\href@noop {} {\bibfield  {journal} {\bibinfo  {journal} {Nature Physics}\ }
  (\bibinfo {year} {2014}{\natexlab{a}})}\BibitemShut {NoStop}%
\bibitem [{\citenamefont {Ashcroft}\ and\ \citenamefont
  {Mermin}(1976)}]{Ashcroft:1976ud}%
  \BibitemOpen
  \bibfield  {author} {\bibinfo {author} {\bibfnamefont {N.~W.}\ \bibnamefont
  {Ashcroft}}\ and\ \bibinfo {author} {\bibfnamefont {N.~D.}\ \bibnamefont
  {Mermin}},\ }\href@noop {} {\emph {\bibinfo {title} {{Solid state
  physics}}}}\ (\bibinfo  {publisher} {Thomson Brooks/Cole},\ \bibinfo {year}
  {1976})\BibitemShut {NoStop}%
\bibitem [{\citenamefont {Kittel}(2004)}]{kittel2004introduction}%
  \BibitemOpen
  \bibfield  {author} {\bibinfo {author} {\bibfnamefont {C.}~\bibnamefont
  {Kittel}},\ }\href@noop {} {\emph {\bibinfo {title} {Introduction to Solid
  State Physics}}}\ (\bibinfo  {publisher} {Wiley},\ \bibinfo {year}
  {2004})\BibitemShut {NoStop}%
\bibitem [{\citenamefont {Feng}\ \emph {et~al.}(1985)\citenamefont {Feng},
  \citenamefont {Thorpe},\ and\ \citenamefont {Garboczi}}]{Feng:1985vr}%
  \BibitemOpen
  \bibfield  {author} {\bibinfo {author} {\bibfnamefont {S.}~\bibnamefont
  {Feng}}, \bibinfo {author} {\bibfnamefont {M.~F.}\ \bibnamefont {Thorpe}}, \
  and\ \bibinfo {author} {\bibfnamefont {E.}~\bibnamefont {Garboczi}},\
  }\href@noop {} {\bibfield  {journal} {\bibinfo  {journal} {Phys. Rev. B}\
  }\textbf {\bibinfo {volume} {31}},\ \bibinfo {pages} {276} (\bibinfo {year}
  {1985})}\BibitemShut {NoStop}%
\bibitem [{\citenamefont {Greaves}\ \emph {et~al.}(2011)\citenamefont
  {Greaves}, \citenamefont {Greer}, \citenamefont {Lakes},\ and\ \citenamefont
  {Rouxel}}]{Greaves:2011ku}%
  \BibitemOpen
  \bibfield  {author} {\bibinfo {author} {\bibfnamefont {G.~N.}\ \bibnamefont
  {Greaves}}, \bibinfo {author} {\bibfnamefont {A.~L.}\ \bibnamefont {Greer}},
  \bibinfo {author} {\bibfnamefont {R.~S.}\ \bibnamefont {Lakes}}, \ and\
  \bibinfo {author} {\bibfnamefont {T.}~\bibnamefont {Rouxel}},\ }\href@noop {}
  {\bibfield  {journal} {\bibinfo  {journal} {Nat Mater}\ }\textbf {\bibinfo
  {volume} {10}},\ \bibinfo {pages} {823} (\bibinfo {year} {2011})}\BibitemShut
  {NoStop}%
\bibitem [{\citenamefont {Witten}(2007)}]{Witten:2007cq}%
  \BibitemOpen
  \bibfield  {author} {\bibinfo {author} {\bibfnamefont {T.}~\bibnamefont
  {Witten}},\ }\href@noop {} {\bibfield  {journal} {\bibinfo  {journal}
  {Reviews of Modern Physics}\ }\textbf {\bibinfo {volume} {79}},\ \bibinfo
  {pages} {643} (\bibinfo {year} {2007})}\BibitemShut {NoStop}%
\bibitem [{\citenamefont {Mahadevan}\ and\ \citenamefont
  {Rica}(2005)}]{Mahadevan:2005hr}%
  \BibitemOpen
  \bibfield  {author} {\bibinfo {author} {\bibfnamefont {L.}~\bibnamefont
  {Mahadevan}}\ and\ \bibinfo {author} {\bibfnamefont {S.}~\bibnamefont
  {Rica}},\ }\href@noop {} {\bibfield  {journal} {\bibinfo  {journal}
  {Science}\ }\textbf {\bibinfo {volume} {307}},\ \bibinfo {pages} {1740}
  (\bibinfo {year} {2005})}\BibitemShut {NoStop}%
\bibitem [{\citenamefont {Castle}\ \emph {et~al.}(2014)\citenamefont {Castle},
  \citenamefont {Cho}, \citenamefont {Gong}, \citenamefont {Jung},
  \citenamefont {Sussman}, \citenamefont {Yang},\ and\ \citenamefont
  {Kamien}}]{Castle:2014jg}%
  \BibitemOpen
  \bibfield  {author} {\bibinfo {author} {\bibfnamefont {T.}~\bibnamefont
  {Castle}}, \bibinfo {author} {\bibfnamefont {Y.}~\bibnamefont {Cho}},
  \bibinfo {author} {\bibfnamefont {X.}~\bibnamefont {Gong}}, \bibinfo {author}
  {\bibfnamefont {E.}~\bibnamefont {Jung}}, \bibinfo {author} {\bibfnamefont
  {D.~M.}\ \bibnamefont {Sussman}}, \bibinfo {author} {\bibfnamefont
  {S.}~\bibnamefont {Yang}}, \ and\ \bibinfo {author} {\bibfnamefont {R.~D.}\
  \bibnamefont {Kamien}},\ }\href@noop {} {\bibfield  {journal} {\bibinfo
  {journal} {Phys. Rev. Lett.}\ }\textbf {\bibinfo {volume} {113}},\ \bibinfo
  {pages} {245502} (\bibinfo {year} {2014})}\BibitemShut {NoStop}%
\bibitem [{\citenamefont {O'Hern}\ \emph {et~al.}(2003)\citenamefont {O'Hern},
  \citenamefont {Silbert}, \citenamefont {Liu},\ and\ \citenamefont
  {Nagel}}]{OHern:2003vq}%
  \BibitemOpen
  \bibfield  {author} {\bibinfo {author} {\bibfnamefont {C.~S.}\ \bibnamefont
  {O'Hern}}, \bibinfo {author} {\bibfnamefont {L.~E.}\ \bibnamefont {Silbert}},
  \bibinfo {author} {\bibfnamefont {A.~J.}\ \bibnamefont {Liu}}, \ and\
  \bibinfo {author} {\bibfnamefont {S.~R.}\ \bibnamefont {Nagel}},\ }\href@noop
  {} {\bibfield  {journal} {\bibinfo  {journal} {Phys. Rev. E}\ }\textbf
  {\bibinfo {volume} {68}},\ \bibinfo {pages} {011306} (\bibinfo {year}
  {2003})}\BibitemShut {NoStop}%
\bibitem [{\citenamefont {Liu}\ and\ \citenamefont {Nagel}(2010)}]{Liu:2010jx}%
  \BibitemOpen
  \bibfield  {author} {\bibinfo {author} {\bibfnamefont {A.~J.}\ \bibnamefont
  {Liu}}\ and\ \bibinfo {author} {\bibfnamefont {S.~R.}\ \bibnamefont
  {Nagel}},\ }\href@noop {} {\bibfield  {journal} {\bibinfo  {journal} {Annu.
  Rev. Condens. Matter Phys.}\ }\textbf {\bibinfo {volume} {1}},\ \bibinfo
  {pages} {347} (\bibinfo {year} {2010})}\BibitemShut {NoStop}%
\bibitem [{\citenamefont {Wyart}\ \emph {et~al.}(2005)\citenamefont {Wyart},
  \citenamefont {Silbert}, \citenamefont {Nagel},\ and\ \citenamefont
  {Witten}}]{Wyart:2005jna}%
  \BibitemOpen
  \bibfield  {author} {\bibinfo {author} {\bibfnamefont {M.}~\bibnamefont
  {Wyart}}, \bibinfo {author} {\bibfnamefont {L.~E.}\ \bibnamefont {Silbert}},
  \bibinfo {author} {\bibfnamefont {S.~R.}\ \bibnamefont {Nagel}}, \ and\
  \bibinfo {author} {\bibfnamefont {T.~A.}\ \bibnamefont {Witten}},\
  }\href@noop {} {\bibfield  {journal} {\bibinfo  {journal} {Phys. Rev. E}\
  }\textbf {\bibinfo {volume} {72}},\ \bibinfo {pages} {051306} (\bibinfo
  {year} {2005})}\BibitemShut {NoStop}%
\bibitem [{\citenamefont {Goodrich}\ \emph {et~al.}(2012)\citenamefont
  {Goodrich}, \citenamefont {Liu},\ and\ \citenamefont
  {Nagel}}]{Goodrich:2012ck}%
  \BibitemOpen
  \bibfield  {author} {\bibinfo {author} {\bibfnamefont {C.~P.}\ \bibnamefont
  {Goodrich}}, \bibinfo {author} {\bibfnamefont {A.~J.}\ \bibnamefont {Liu}}, \
  and\ \bibinfo {author} {\bibfnamefont {S.~R.}\ \bibnamefont {Nagel}},\
  }\href@noop {} {\bibfield  {journal} {\bibinfo  {journal} {Phys. Rev. Lett.}\
  }\textbf {\bibinfo {volume} {109}},\ \bibinfo {pages} {095704} (\bibinfo
  {year} {2012})}\BibitemShut {NoStop}%
\bibitem [{\citenamefont {Dagois-Bohy}\ \emph {et~al.}(2012)\citenamefont
  {Dagois-Bohy}, \citenamefont {Tighe}, \citenamefont {Simon}, \citenamefont
  {Henkes},\ and\ \citenamefont {van Hecke}}]{DagoisBohy:2012dh}%
  \BibitemOpen
  \bibfield  {author} {\bibinfo {author} {\bibfnamefont {S.}~\bibnamefont
  {Dagois-Bohy}}, \bibinfo {author} {\bibfnamefont {B.}~\bibnamefont {Tighe}},
  \bibinfo {author} {\bibfnamefont {J.}~\bibnamefont {Simon}}, \bibinfo
  {author} {\bibfnamefont {S.}~\bibnamefont {Henkes}}, \ and\ \bibinfo {author}
  {\bibfnamefont {M.}~\bibnamefont {van Hecke}},\ }\href@noop {} {\bibfield
  {journal} {\bibinfo  {journal} {Phys. Rev. Lett.}\ }\textbf {\bibinfo
  {volume} {109}},\ \bibinfo {pages} {095703} (\bibinfo {year}
  {2012})}\BibitemShut {NoStop}%
\bibitem [{\citenamefont {Goodrich}\ \emph
  {et~al.}(2014{\natexlab{b}})\citenamefont {Goodrich}, \citenamefont
  {Dagois-Bohy}, \citenamefont {Tighe}, \citenamefont {van Hecke},
  \citenamefont {Liu},\ and\ \citenamefont {Nagel}}]{Goodrich:2014iu}%
  \BibitemOpen
  \bibfield  {author} {\bibinfo {author} {\bibfnamefont {C.~P.}\ \bibnamefont
  {Goodrich}}, \bibinfo {author} {\bibfnamefont {S.}~\bibnamefont
  {Dagois-Bohy}}, \bibinfo {author} {\bibfnamefont {B.~P.}\ \bibnamefont
  {Tighe}}, \bibinfo {author} {\bibfnamefont {M.}~\bibnamefont {van Hecke}},
  \bibinfo {author} {\bibfnamefont {A.~J.}\ \bibnamefont {Liu}}, \ and\
  \bibinfo {author} {\bibfnamefont {S.~R.}\ \bibnamefont {Nagel}},\ }\href@noop
  {} {\bibfield  {journal} {\bibinfo  {journal} {Phys. Rev. E}\ }\textbf
  {\bibinfo {volume} {90}},\ \bibinfo {pages} {022138} (\bibinfo {year}
  {2014}{\natexlab{b}})}\BibitemShut {NoStop}%
\bibitem [{\citenamefont {Ellenbroek}\ \emph {et~al.}(2009)\citenamefont
  {Ellenbroek}, \citenamefont {Zeravcic}, \citenamefont {van Saarloos},\ and\
  \citenamefont {van Hecke}}]{Ellenbroek:2009to}%
  \BibitemOpen
  \bibfield  {author} {\bibinfo {author} {\bibfnamefont {W.~G.}\ \bibnamefont
  {Ellenbroek}}, \bibinfo {author} {\bibfnamefont {Z.}~\bibnamefont
  {Zeravcic}}, \bibinfo {author} {\bibfnamefont {W.}~\bibnamefont {van
  Saarloos}}, \ and\ \bibinfo {author} {\bibfnamefont {M.}~\bibnamefont {van
  Hecke}},\ }\href@noop {} {\bibfield  {journal} {\bibinfo  {journal} {EPL}\
  }\textbf {\bibinfo {volume} {87}},\ \bibinfo {pages} {34004} (\bibinfo {year}
  {2009})}\BibitemShut {NoStop}%
\bibitem [{\citenamefont {Ellenbroek}\ \emph {et~al.}(2014)\citenamefont
  {Ellenbroek}, \citenamefont {Hagh}, \citenamefont {Kumar}, \citenamefont
  {Thorpe},\ and\ \citenamefont {van Hecke}}]{Ellenbroek:2014uh}%
  \BibitemOpen
  \bibfield  {author} {\bibinfo {author} {\bibfnamefont {W.~G.}\ \bibnamefont
  {Ellenbroek}}, \bibinfo {author} {\bibfnamefont {V.~F.}\ \bibnamefont
  {Hagh}}, \bibinfo {author} {\bibfnamefont {A.}~\bibnamefont {Kumar}},
  \bibinfo {author} {\bibfnamefont {M.~F.}\ \bibnamefont {Thorpe}}, \ and\
  \bibinfo {author} {\bibfnamefont {M.}~\bibnamefont {van Hecke}},\ }\href@noop
  {} {\bibfield  {journal} {\bibinfo  {journal} {arXiv}\ } (\bibinfo {year}
  {2014})},\ \Eprint {http://arxiv.org/abs/1412.0273v1} {1412.0273v1}
  \BibitemShut {NoStop}%
\bibitem [{\citenamefont {Levine}(2001)}]{Levine2001}%
  \BibitemOpen
  \bibfield  {author} {\bibinfo {author} {\bibfnamefont {D.}~\bibnamefont
  {Levine}},\ }in\ \href@noop {} {\emph {\bibinfo {booktitle} {Jamming and
  Rheology: Constrained Dynamics on Microscopic and Macroscopic Scales}}},\
  \bibinfo {editor} {edited by\ \bibinfo {editor} {\bibfnamefont
  {A.}~\bibnamefont {Liu}}\ and\ \bibinfo {editor} {\bibfnamefont
  {S.}~\bibnamefont {Nagel}}}\ (\bibinfo  {publisher} {Taylor \& Francis},\
  \bibinfo {address} {London},\ \bibinfo {year} {2001})\BibitemShut {NoStop}%
\bibitem [{\citenamefont {Driscoll}\ \emph {et~al.}(2015)\citenamefont
  {Driscoll}, \citenamefont {Chen}, \citenamefont {Beuman}, \citenamefont
  {Ulrich}, \citenamefont {Nagel},\ and\ \citenamefont
  {Vitelli}}]{Driscoll:2015vf}%
  \BibitemOpen
  \bibfield  {author} {\bibinfo {author} {\bibfnamefont {M.~M.}\ \bibnamefont
  {Driscoll}}, \bibinfo {author} {\bibfnamefont {B.~G.-G.}\ \bibnamefont
  {Chen}}, \bibinfo {author} {\bibfnamefont {T.~H.}\ \bibnamefont {Beuman}},
  \bibinfo {author} {\bibfnamefont {S.}~\bibnamefont {Ulrich}}, \bibinfo
  {author} {\bibfnamefont {S.~R.}\ \bibnamefont {Nagel}}, \ and\ \bibinfo
  {author} {\bibfnamefont {V.}~\bibnamefont {Vitelli}},\ }\href@noop {}
  {\bibfield  {journal} {\bibinfo  {journal} {arXiv}\ } (\bibinfo {year}
  {2015})},\ \Eprint {http://arxiv.org/abs/1501.04227v1} {1501.04227v1}
  \BibitemShut {NoStop}%
\bibitem [{\citenamefont {Dearcangelis}\ \emph {et~al.}(1985)\citenamefont
  {Dearcangelis}, \citenamefont {Redner},\ and\ \citenamefont
  {Coniglio}}]{DEARCANGELIS:1985uh}%
  \BibitemOpen
  \bibfield  {author} {\bibinfo {author} {\bibfnamefont {L.}~\bibnamefont
  {Dearcangelis}}, \bibinfo {author} {\bibfnamefont {S.}~\bibnamefont
  {Redner}}, \ and\ \bibinfo {author} {\bibfnamefont {A.}~\bibnamefont
  {Coniglio}},\ }\href@noop {} {\bibfield  {journal} {\bibinfo  {journal}
  {Phys. Rev. B}\ }\textbf {\bibinfo {volume} {31}},\ \bibinfo {pages} {4725}
  (\bibinfo {year} {1985})}\BibitemShut {NoStop}%
\bibitem [{\citenamefont {Dearcangelis}\ \emph {et~al.}(1986)\citenamefont
  {Dearcangelis}, \citenamefont {Redner},\ and\ \citenamefont
  {Coniglio}}]{DEARCANGELIS:1986vk}%
  \BibitemOpen
  \bibfield  {author} {\bibinfo {author} {\bibfnamefont {L.}~\bibnamefont
  {Dearcangelis}}, \bibinfo {author} {\bibfnamefont {S.}~\bibnamefont
  {Redner}}, \ and\ \bibinfo {author} {\bibfnamefont {A.}~\bibnamefont
  {Coniglio}},\ }\href@noop {} {\bibfield  {journal} {\bibinfo  {journal}
  {Phys. Rev. B}\ }\textbf {\bibinfo {volume} {34}},\ \bibinfo {pages} {4656}
  (\bibinfo {year} {1986})}\BibitemShut {NoStop}%
\bibitem [{\citenamefont {Keyak}\ \emph {et~al.}(2013)\citenamefont {Keyak},
  \citenamefont {Sigurdsson}, \citenamefont {Karlsdottir}, \citenamefont
  {Oskarsdottir}, \citenamefont {Sigmarsdottir}, \citenamefont {Kornak},
  \citenamefont {Harris}, \citenamefont {Sigurdsson}, \citenamefont {Jonsson},
  \citenamefont {Siggeirsdottir}, \citenamefont {Eiriksdottir}, \citenamefont
  {Gudnason},\ and\ \citenamefont {Lang}}]{Keyak:2013ga}%
  \BibitemOpen
  \bibfield  {author} {\bibinfo {author} {\bibfnamefont {J.~H.}\ \bibnamefont
  {Keyak}}, \bibinfo {author} {\bibfnamefont {S.}~\bibnamefont {Sigurdsson}},
  \bibinfo {author} {\bibfnamefont {G.~S.}\ \bibnamefont {Karlsdottir}},
  \bibinfo {author} {\bibfnamefont {D.}~\bibnamefont {Oskarsdottir}}, \bibinfo
  {author} {\bibfnamefont {A.}~\bibnamefont {Sigmarsdottir}}, \bibinfo {author}
  {\bibfnamefont {J.}~\bibnamefont {Kornak}}, \bibinfo {author} {\bibfnamefont
  {T.~B.}\ \bibnamefont {Harris}}, \bibinfo {author} {\bibfnamefont
  {G.}~\bibnamefont {Sigurdsson}}, \bibinfo {author} {\bibfnamefont {B.~Y.}\
  \bibnamefont {Jonsson}}, \bibinfo {author} {\bibfnamefont {K.}~\bibnamefont
  {Siggeirsdottir}}, \bibinfo {author} {\bibfnamefont {G.}~\bibnamefont
  {Eiriksdottir}}, \bibinfo {author} {\bibfnamefont {V.}~\bibnamefont
  {Gudnason}}, \ and\ \bibinfo {author} {\bibfnamefont {T.~F.}\ \bibnamefont
  {Lang}},\ }\href@noop {} {\bibfield  {journal} {\bibinfo  {journal} {Bone}\
  }\textbf {\bibinfo {volume} {57}},\ \bibinfo {pages} {18} (\bibinfo {year}
  {2013})}\BibitemShut {NoStop}%
\bibitem [{\citenamefont {Lenz}\ \emph {et~al.}(2012)\citenamefont {Lenz},
  \citenamefont {Thoresen}, \citenamefont {Gardel},\ and\ \citenamefont
  {Dinner}}]{Lenz:2012df}%
  \BibitemOpen
  \bibfield  {author} {\bibinfo {author} {\bibfnamefont {M.}~\bibnamefont
  {Lenz}}, \bibinfo {author} {\bibfnamefont {T.}~\bibnamefont {Thoresen}},
  \bibinfo {author} {\bibfnamefont {M.~L.}\ \bibnamefont {Gardel}}, \ and\
  \bibinfo {author} {\bibfnamefont {A.~R.}\ \bibnamefont {Dinner}},\
  }\href@noop {} {\bibfield  {journal} {\bibinfo  {journal} {Phys. Rev. Lett.}\
  }\textbf {\bibinfo {volume} {108}},\ \bibinfo {pages} {238107} (\bibinfo
  {year} {2012})}\BibitemShut {NoStop}%
\end{thebibliography}


\end{document}